# DEVELOPMENT AND VALIDATION OF A TEACHING PRACTICE SCALE (TISS) FOR INSTRUCTORS OF INTRODUCTORY STATISTICS AT THE COLLEGE LEVEL

HASSAD, Rossi
Mercy College
New York, USA

This study examined the teaching practices of 227 college instructors of introductory statistics (from the health and behavioral sciences). Using primarily multidimensional scaling (MDS) techniques, a two-dimensional, 10-item teaching practice scale, TISS (**T**eaching of **I**ntroductory **S**tatistics **S**cale), was developed and validated. The two dimensions (subscales) were characterized as constructivist, and behaviorist, and are orthogonal to each other. Criterion validity of this scale was established in relation to instructors' attitude toward teaching, and acceptable levels of reliability were obtained. A significantly higher level of behaviorist practice (less reform-oriented) was reported by instructors from the USA, and instructors with academic degrees in mathematics and engineering. This new scale (TISS) will allow us to empirically assess and describe the pedagogical approach (teaching practice) of instructors of introductory statistics. Further research is required in order to be conclusive about the structural and psychometric properties of this scale.

INTRODUCTION

The ability to critically evaluate research findings (often expressed in statistical jargon) is an essential skill for practitioners and students in the traditional evidence-based disciplines such as the health and behavioral sciences (Belar, 2003). Toward achieving this competency, undergraduate students in these disciplines are generally required to take introductory statistics as a core course, and there is a consensus among educators that the goal of this course should be to facilitate statistical literacy and thinking through active learning strategies, by emphasizing concepts and applications rather than mathematical procedures (Franklin & Garfield, 2006). Also underpinning this pedagogical approach is the realization that for the majority of these students, the introductory course will be their only formal exposure to statistics (Moore, 1998).

For over a decade there has been emphasis on reform-oriented teaching at the college level, fueled by a consensus among educators that traditional curricular material and pedagogical strategies have not been effective in promoting statistical literacy and thinking (Cobb, 1992; Delmas et al., 2006; Garfield et al., 2002). In spite of these reform efforts focused on course content, pedagogy, assessment, and integration of technology, research continues to show that students are emerging with a lack of understanding of core concepts (Delmas et al., 2006). Such evidence has raised concerns about instructors' level of awareness, understanding, and appropriate use of active learning strategies (Hassad, 2007). Also, empirical information on what core strategies underlie reform-oriented teaching of introductory statistics is lacking (Garfield et al., 2002), and this is a major impediment to characterizing teaching practice, and assessing the effectiveness of reform-oriented teaching compared to the traditional pedagogical approach.

OBJECTIVE

The objective of this study was to develop and validate a scale (instrument) to empirically assess and describe the pedagogical approach (teaching practice) of instructors of introductory statistics in the health and behavioral sciences, at the college level. Such a scale can be used to characterize teaching practice, toward identifying individual strengths and weaknesses regarding reform-oriented (constructivist or concept-based) teaching of introductory statistics. More importantly, this instrument will allow us to determine what learning outcomes result from the different practice orientations. This study also identifies elements of the undergraduate curriculum that should be emphasized so as to facilitate more effective preparation of our students for the workplace and postgraduate studies.

THEORETICAL BASIS OF REFORM-ORIENTED PEDAGOGY

Reform in this context, represents a shift in pedagogical philosophy from the behaviorist to the constructivist (Caprio, 1994; Trigwell & Prosser, 2004). In general, behaviorist-oriented instructors tend to be more preoccupied with subject content, and the transmission of information (passive student and a top-down approach) whereas constructivist-oriented instructors are more student-centered, and aim to facilitate the construction of knowledge and meaning by the student (active student and a bottom-up approach) (Trigwell & Prosser, 2004). In the constructivist context, the instructor seeks to engender meaningful (deep and conceptual) learning, that is, the ability to know "what to do and why" (Skemp, 1987, p.9) unlike teachers with the behaviorist orientation who foster rote learning (and surface knowledge) by focusing on discrete and compartmentalized knowledge and skills.

REFORM-ORIENTED PEDAGOGY AND STATISTICAL LITERACY

The reform-oriented (concept-based or constructivist) approach to teaching introductory statistics) is generally operationalized as a set of active learning strategies intended to facilitate statistical literacy. Such active learning strategies include projects, group discussions, data collection, hands-on computer data analysis, critiquing of research articles, report writing, oral presentations, and the use of real-world data. Statistical literacy (thinking and reasoning) refers to the ability to understand, critically evaluate, and use statistical information and data-based arguments (Gal, 2000; Garfield et al., 2002). The GAISE (**G**uidelines for **A**ssessment and **I**nstruction in **S**tatistics **E**ducation) report (Franklin & Garfield, 2006) which serves as a universal blueprint for reform-oriented teaching of introductory statistics, recommends the following:

1. Emphasize statistical literacy and develop statistical thinking;
2. Use real data;
3. Stress conceptual understanding rather than mere knowledge of procedures;
4. Foster active learning in the classroom;
5. Use technology for developing conceptual understanding and analyzing data;
6. Use assessments to improve and evaluate student learning.

METHODOLOGY

*STUDY DESIGN, SUBJECTS, AND SAMPLE METHODOLOGY*

The development of this teaching practice scale was one component of an initial exploratory cross-sectional study which concurrently developed and validated an attitude scale (Hassad & Coxon, 2007) for instructors of undergraduate introductory statistics. The subjects were a purposive (maximum variation) sample of 227 instructors from the health and behavioral sciences at four-year regionally accredited academic institutions in the USA (and the foreign equivalent).

*DEVELOPMENT OF THE TEACHING PRACTICE ITEMS*

Teaching practice was conceptualized as a continuum, that is, high reform (concept-based or constructivist) to low reform (traditional or behaviorist). The scale content was guided by the GAISE (**G**uidelines for **A**ssessment and **I**nstruction in **S**tatistics **E**ducation) report on introductory statistics (Franklin & Garfield, 2006), as well as the Cobb report (Cobb, 1992). The initial set of items was culled from related studies, and formulated in consultation with pioneer statistics educators. Item analysis was performed by a multidisciplinary team of college instructors, focusing on content and face validity, salience, clarity, theoretical and empirical relevance, and redundancy. A pilot test was conducted via email, and this resulted in a final set of 10 practice items (behaviorist and constructivist) on a frequency of use scale of 1 (never) through 5 (always).

*RECRUITMENT OF SUBJECTS & DATA COLLECTION*

In order to represent the range of teaching practice, recruitment involved targeting instructors at colleges where statistics educators active in the reform movement were employed. Instructors were also targeted based on their publications and course outlines. Faculty charcaterized as traditional or beahviorist were equally targeted. Additional contacts were obtained from journal articles, conference proceedings, and listservs. The questionnaire was programmed in Hyper Text Markup Language (HTML), and three emails (an invitation to participate, a reminder, and a last call to participate) were sent (one week apart), with an online link to the questionnaire. Online informed consent was obtained, and data collection took place between August and October of 2005. The completed questionnaires were checked for redundant or duplicate submissions, and as an incentive for participation, three one-hundred dollar awards were raffled.

*DATA ANALYSIS (MDS)*

The structural (underlying dimensions) and psychometric (reliability and validity) properties of the teaching practice data were examined using primarily multidimensional scaling (MDS) techniques. Also, selected factors were explored as correlates of teaching practice. The behaviorist items were reverse-coded to obtain meaningful scores. MDS seeks to reduce and organize the data to achieve a spatial representation (geometric map) of the latent structure that underlies the relationships among the items (Coxon, 1982; Kruskal & Wish, 1978). MDS has both metric (linear transformation) and non-metric (ordinal transformation) variants, unlike factor analysis which requires the assumptions of metric data, and linear relationships (Coxon, 1982). These teaching practice data are ordinal (obtained on a Likert-type scale), and hence non-metric, therefore MDS is suitable for this study. The input information for MDS is a numerical measure of distance, indicating how similar each item is to the others. Both metric (MRSCAL) and non-metric (MINISSA) MDS (Coxon, 1982) were performed, with Pearson's correlation coefficient (based on the interval properties of the data) and Kendall's tau (based on the rank order of the data) as measures of similarity.

*INTERPRETATION OF THE MDS MAPS (CONFIGURATIONS)*

Interpretation involved identifying and assigning meanings to patterns or regions (clusters of items), and for this, a two-dimensional configuration is recommended (Kruskal & Wish, 1978). Also, hierarchical cluster analysis was used to guide the identification of patterns within the spatial maps (Coxon, 1982), which were rotated to simple structure (Kruskal & Wish, 1978). The adequacy of the MDS solutions was evaluated by the stress and the coefficient of determination (R-squared) values. Both values are measures of goodness of fit between the input data and the MDS model (Coxon, 1982). The stability of the solutions was assessed using the guideline of at least $4k + 1$ objects (items) for a *k*-dimensional solution, as well as consistency across all the MDS maps (Kruskal & Wish, 1978).

*RELIABILITY AND VALIDITY ANALYSIS*

Cronbach's alpha (Cronbach, 1951) which quantifies the degree of internal consistency (reliability) of a set of items, was calculated for each subscale, as well as the overall scale. In general, a Cronbach's alpha of at least .7 is the criterion used to establish an acceptable level of reliability. However, the recommended minimum Cronbach's alpha for exploratory studies is .6 (Nunnally, 1978; Robinson, Shaver, & Wrightsman, 1991). Validity is a multidimensional concept (more appropriately labeled construct validity), and refers to whether the scale measures the construct (teaching practice) as theorized (Cronbach & Meehl, 1955). Criterion validity is generally considered the core dimension of construct validity (Muldoon et al., 1998), and is reported herein. According to Cronbach & Meehl (1955), in order to establish criterion validity we must show that the construct (being assessed) relates to another construct (the criterion) in a theoretically predictable way. In this regard, the attitude-practice relationship was explored with the expectation that attitude scores will meaningfully differentiate between high and low-reform practice instructors. An overall teaching practice score was calculated for each subject based on the sum of the ten practice item scores (with the behaviorist items reverse-coded). The maximum possible practice score was therefore 50, and respondents in the highest quartile were classified as high-reform instructors, whereas those in the lowest quartile were labeled low-reform instructors. The attitude scale, **FATS** (**F**aculty **A**ttitudes **T**oward **S**tatistics) (Hassad & Coxon, 2007) was used. It was validated, and consists of five subscales with a total of 25 items, and an overall alpha of .89. The mean scale score was computed for each subject (with a maximum possible score of 5). Additionally, multiple regression analysis of teaching practice score on attitude subscale scores was performed to determine the extent to which attitude can predict teaching practice in this context. Statistical significance was determined based on an alpha level of .05, and Bonferroni adjustment for multiple comparison testing was performed where applicable.

RESULTS & DISCUSSION

*RESPONDENTS' BACKGROUND CHARACTERISTICS*
There were 227 participants, and of the 222 who provided country information, 165 (74%) were from the USA, and 57 (26%) from international locations (primarily the UK, Netherlands, Canada, and Australia). They represented 24 countries and 133 academic institutions. The median age category and duration of teaching were 41 to 50 years, and 10 years respectively. The majority were male, 139 (61%), and from the USA sub-sample, 135 (82%) identified as Caucasian. There were 94 (41%) instructors from the health sciences, 102 (45%) from the behavioral sciences, and 31(14%) who taught both in the health and behavioral sciences. The modal category for academic degree concentration was statistics, 92 (41%), followed by psychology/social/behavioral sciences, 71(31%). The academic specialization least reported was mathematics/engineering, 17 (8%).

*MULTIDIMENSIONAL SCALING (MDS) OF THE TEACHING PRACTICE ITEMS*
In order to identify the latent structure underlying the interrelationships among the ten (10) practice items, both metric (MRSCAL), and non-metric MDS (MINISSA) procedures were conducted, and 1 to 3-dimensional maps were generated. The two-dimensional maps were the most meaningful and interpretable, and the best fit (Figure 1) was obtained with non-metric MDS using Pearson's correlation coefficient as the input measure of similarity. This map (Figure 1) reveals two distinct clusters, separating the practice items as theorized, that is, behaviorist (low-reform practice), and constructivist (high-reform practice).

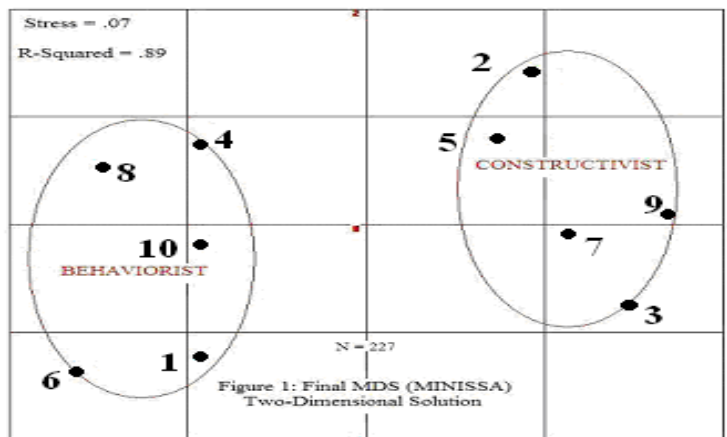

Figure 1: Final MDS (MINISSA) Two-Dimensional Solution

Refer to Table 1 for Number Labels

Also, this solution or map fits the data very well, with a normalized stress value (residual sum of squares) of .07. Stress values closer to zero represent a better fit. In this case, the stress value is more than two times smaller than stress based on simulation approximation to random data (Spence, 1979). The amount of variance within the data that is accounted for by this two-dimensional solution, is 89% (R-squared), suggesting a very good fit. Additionally, the stability of the solution is in keeping with the empirical guideline of at least $4k + 1$ objects for a $k$-dimensional solution with non-metric scaling (Kruskal & Wish, 1978).

| Table 1: Teaching of Introductory Statistics Scale (TISS) (α = .6) | | | | | |
|---|---|---|---|---|---|
| **Practice Items** | **Never** **1** | **Rarely** **2** | **Sometimes** **3** | **Usually** **4** | **Always** **5** |
| 1. I emphasize rules and formulas as a basis for subsequent learning.* | | | | | |
| 2. I integrate statistics with other subjects. | | | | | |
| 3. Students use a computer program to explore and analyze data. | | | | | |
| 4. I assign homework primarily from the textbook.* | | | | | |
| 5. Critiquing of research articles is a core learning activity. | | | | | |
| 6. The mathematical underpinning of each statistical test is emphasized.* | | | | | |
| 7. I use real-life data for class demonstrations and assignments. | | | | | |
| 8. I require that students adhere to procedures in the textbook.* | | | | | |
| 9. Assessment includes written reports of data analysis. | | | | | |
| 10. I assign drill and practice exercises (mathematical) for each topic.* | | | | | |
| *These items must be reverse-coded for the overall teaching practice score, so that higher values reflect higher levels of reform-oriented (concept-based or constructivist) practice. **Behaviorist subscale**: Items 1,4,6,8,10 (α =.61), **Constructivist subscale:** Items 2,3,5,7,9 (α = .66) | | | | | |

*RELIABILITY ANALYSIS & INTER-SUBSCALE CORRELATION*

The Cronbach's alpha (**α**) of the overall scale is .6, an acceptable level of reliability for exploratory studies (Nunnally, 1978; Robinson, Shaver, & Wrightsman, 1991). Furthermore, each subscale (behaviorist: **α** = .61, constructivist: **α** =.66) is more internally consistent than the overall scale, and this could support the finding that two dimensions underlie teaching practice (Yu, 2001). Deletion of any item did not appreciably improve reliability, and the item-total correlations are .3 or higher, indicating that each item is meaningful to the scale (Nunnally & Bernstein, 1994). The behaviorist and constructivist subscales are almost orthogonal of each other (Pearson's r = -.06, df = 217, ns), and hence can be considered independent dimensions of teaching practice. Similar findings (from different disciplines) were reported by Woolley, Benjamin and Woolley (2004), and Handal (2002) who obtained correlation coefficients of -0.232 and -.011 respectively. The absence of correlation between the two subscales can be viewed as strong evidence that the behaviorist subscale (formed by the reverse-coded items) is a meaningful and separable dimension of teaching practice, and not a "method" or artifactual factor (Hall et al., 2002).

*CRITERION VALIDITY*

Criterion validity was based on theoretically predictable differences between high and low-reform instructors with respect to attitude scores (measured using the FATS scale – Faculty Attitudes Toward Statistics). According to attitude theory, and the attitude-behavior relationship (Wallace et al., 2005), high-reform instructors (those in the highest quartile of practice score) should possess more favorable attitude (higher attitude scores) toward constructivist teaching than low-reform instructors (those in the lowest quartile of practice score). As shown in Table 2, high-reform instructors, on average, did report higher scores (more favorable disposition toward constructivist pedagogy) on the overall attitude scale, as well as each of the five subscales, and all but "perceived difficulty" (ease of use) were statistically significant. Multiple regression analysis was next conducted to determine the predictive value of the 5-factor attitude scale in relation to teaching practice.

**Table 2: Comparison of Overall Attitude and Subscale Scores by Teaching Practice Categorization**

| Attitude Subscales | Teaching Practice Score [a] | N | Mean | Std. Dev. | t | t-test | Mann-Whitney |
|---|---|---|---|---|---|---|---|
| Perceived Usefulness | Lowest Quartile | 74 | 3.84 | .63 | 5.57 | .001 | .001 |
| | Highest Quartile | 59 | 4.41 | .50 | | | |
| Intention | Lowest Quartile | 74 | 3.57 | .77 | 6.90 | .001 | .001 |
| | Highest Quartile | 59 | 4.36 | .55 | | | |
| Personal Teaching Efficacy | Lowest Quartile | 74 | 3.71 | .66 | 7.47 | .001 | .001 |
| | Highest Quartile | 59 | 4.43 | .45 | | | |
| Avoidance-Approach | Lowest Quartile | 73 | 3.82 | .70 | 4.48 | .001 | .001 |
| | Highest Quartile | 59 | 4.32 | .52 | | | |
| Perceived Difficulty | Lowest Quartile | 73 | 2.93 | .76 | 1.16 | .248 | .211 |
| | Highest Quartile | 59 | 3.10 | .92 | | | |
| **Overall Attitude (Total)** | Lowest Quartile | 72 | 3.65 | .47 | 8.26 | .001 | .001 |
| | Highest Quartile | 59 | 4.23 | .33 | | | |

a. Lowest Quartile = Low Reform Instructors
Highest Quartile = High Reform Instructors

*MULTIPLE REGRESSION ANALYSIS OF TEACHING PRACTICE ON ATTITUDE*

Teaching practice score (the dependent variable) was regressed on the five attitude subscale scores (the independent variables). The overall model (Table 3) was statistically significant, and explained 28% (see adjusted $R^2$) of the variance in teaching practice, which is consistent with major attitude-behavior research (Armitage & Conner, 2001). Intention (one component of attitude) was the strongest predictor of practice, a finding that is both theoretically and empirically well-supported (Wallace et al., 2005; Armitage & Conner, 2001).

**Table 3: Multiple Regression Analysis of Overall Teaching Practice Score on Attitude Subscale Scores**

| Independent Variables (subscales)[a] | Unstandardized Coefficients | | Standardized Coefficients | | |
|---|---|---|---|---|---|
| | B | Std. Error | Beta | t | Sig. |
| (Constant) | 17.05 | 2.30 | | 7.42 | .001 |
| Perceived Usefulness | .04 | .61 | .01 | .07 | .946 |
| Intention | 1.58 | .52 | .26 | 3.01 | .003 |
| Personal Teaching Efficacy | 1.62 | .54 | .24 | 3.01 | .003 |
| Avoidance-Approach | 1.46 | .47 | .20 | 3.07 | .002 |
| Perceived Difficulty | -.42 | .35 | -.08 | -1.22 | .225 |

a. Model Significance: F (5, 208) = 17.3, p<.001, Adjusted R-Squared = .28 (28%)

*CORRELATES OF TEACHING PRACTICE SUBSCALE SCORES*

Subscale scores (constructivist and behaviorist) did not vary significantly with respect to gender, age, ethnicity, duration of teaching, teaching area, membership status in professional organizations, degree concentration, and employment status. However, significant differences (p<.05) were noted as follows: Instructors from international locations (Mean = 12, SD = 3.27), reported a lower level of behaviorist practice than those from the USA (Mean = 14, SD = 2.85). Also, those with mathematics and engineering degrees (Mean = 15, SD = 2.88) had the highest level of behaviorist practice compared to those with health sciences degrees (Mean= 12, SD = 3.22) who had the lowest level on this scale.

CONCLUSION AND RECOMMENDATIONS

This exploratory cross-sectional study examined the teaching practices of 227 college instructors of introductory statistics (from the health and behavioral sciences). Using primarily multidimensional scaling (MDS) techniques, a two-dimensional, 10-item teaching practice scale was developed and validated (Table 1). This scale will be referred to as **TISS** (**T**eaching of **I**ntroductory **S**tatistics **S**cale). The two dimensions (subscales) were characterized as **constructivist** (reform-oriented, student-centered, and active learning), and **behaviorist** (instructor-centered, and passive learning), and are orthogonal (not correlated). The absence of correlation between the constructivist and behaviorist subscales, strongly support that these are independent dimensions of teaching practice. The constructivist subscale reflects integration of information from other subjects, use of computers, critiquing of research articles, use of real-world data, and written reports, whereas the behaviorist subscale reflects emphasis on: rules and formulas, mathematical underpinnings, drill and practice exercises, and textbook-centeredness. Moreover, criterion validity of this scale was established in relation to instructors' attitude toward teaching, and acceptable levels of reliability (internal consistency) were obtained for the overall scale as well as both subscales.

Teaching practice was neither exclusively constructivist nor behaviorist, but for most instructors, predominantly one or the other. This eclectic pedagogical approach, in particular, the extent of use of strategies from each practice orientation, is quite likely context-dependent. In this regard, a significantly higher level of behaviorist practice (less reform-oriented) was reported by instructors from the USA (compared to international locations), and instructors with their highest academic degree in either mathematics or engineering compared to those with degrees in health sciences (who reported the lowest level of behaviorist practice). The teaching reform movement discourages behaviorist practices, and promotes constructivist pedagogy, therefore, these differences warrant further exploration. This scale can be used to characterize teaching practice, toward identifying individual strengths and weaknesses regarding constructivist (reform-oriented or concept-based) teaching of introductory statistics. Accordingly, targeted professional development programs can be developed to facilitate and maintain such practice.

This is an initial exploratory cross-sectional study, and further research will be required in order to be conclusive about the structural and psychometric properties of this scale. Furthermore, this study examined internal consistency (reliability), and not test-retest reliability, which should be assessed in order to determine the stability or robustness of the scale. Indeed, this new scale (TISS) will allow us to empirically assess and describe the pedagogical approach (teaching practice) of instructors of introductory statistics in the health and behavioral sciences, at the college level, and determine what learning outcomes result from the different teaching practice orientations. Maybe, we can now better respond to a major concern about introductory statistics education, expressed by Garfield et al. (2002), that is, "no one has yet demonstrated that a particular set of teaching techniques or materials will lead to the desired outcomes".

REFERENCES

Armitage, C. J., & Conner, M. (1999). The theory of planned behaviour: Assessment
 of predictive validity and "perceived control." *British Journal of Social Psychology*, 38, 35-54.
Belar, C. (2003). Training for evidence-based practice. *APA Monitor on Psychology*,
 Volume 34, No. 5., p56.
Caprio, M. (1994). Easing into constructivism. *Journal of College Science Teaching*,
 23(6), pp. 210-212.
Cobb, G. W. (1992). Teaching statistics. In L. Steen (Ed.), Heeding the call for change:
 Suggestions for curricular action, *MAA Notes*, No. 22, 3-43.

Coxon, A.P.M. (1982). *The user's guide to multidimensional scaling*. London: Heinemann Educational Books.

Cronbach, L. J. (1951). Coefficient alpha and the internal structure of tests. *Psychometrika*, 16(3), 297-334.

Cronbach, L.J. & Meehl, P.C. (1955). Construct validity in psychological tests. *Psychological Bulletin*, 52: 281-302.

delmas, R., Garfield, J., Ooms, A., & Chance, B. (2006). Assessing Students' Conceptual Understanding After a First Course in Statistics. Paper presented at the Annual Meeting of *The American Educational Research Association (*AERA) San Francisco, CA.

Franklin, C., & Garfield, J. (2006). Guidelines for Statistics Education Endorsed by ASA Board of Directors. *Amstat News* (Education), Issue No. 348.

Gal, I. (2000). Statistical literacy: Conceptual and instructional issues. In D. Coben, J. O'Donoghue & G. E. Fitzsimons (Eds.), *Perspectives on adults learning mathematics* (pp. 135-150). Boston: Kluwer Academic Publishers.

Garfield, J., Hogg, B., Schau, C., & Whittinghill, D. (2002). First Courses in Statistical Science: The Status of Educational Reform Efforts. *Journal of Statistics Education* [Online], 10(2).

Hall, M.A., Camacho F., Dugan E., & Balkrishnan, R. (2002). Trust in the Medical Profession: Conceptual and Measurement Issues. *Health Serv Res*., 37(5): 1419-39.

Handal, B. (2004). Teachers' instructional beliefs about integrating educational technology. *e-Journal of Instructional Science and Technology,* 17(1).

Hassad, R., & Coxon, A. (2007). Development and Validation of a Scale to Measure Instructors'Attitudes to Concept-Based Teaching of Introductory Statistics in the Health and Behavioral Sciences – *Bulletin of the International Statistical Institute*, August 2007.

Kruskal, J. B., & Wish, M. (1978). *Multidimensional Scaling*. Newbury Park, CA: Sage.

Moore, D. S. (1988). Statistics Among the Liberal Arts. *Journal of the American Statistical Association*, Vol. 93, No. 444, pp. 1253-1259.

Muldoon, M. F., Barger, S., Flory, J., & Manuck, S. (1998). What are quality of life measurements measuring? *British Medical Journal*, Vol. 316, No. 7130, 542 –545.

Nunnally, J. C., & Bernstein, I. H. (1994). *Psychometric theory* (3$^{rd}$ Ed.). New York: McGraw-Hill.

Nunnally, J.C. (1978). Psychometric Theory (2nd ed.). New York: McGraw Hill.

Robinson, J. P., Shaver, P. R., & Wrightsman, L. S. (Eds.). (1991). *Measures of personality an social psychological attitudes.* San Diego: Academic Press.

Skemp, R.R (1987). *The psychology of learning mathematics*. NJ Hillsdale: Lawrence

Spence, I. (1979) A simple approximation for random rankings stress values. *Multivariate Behavioral Research, 14*, 355-365.

Trigwell, K., & Prosser, M. (2004). Development and use of the Approaches to Teaching Inventory. *Educational Psychology Review*, 16(4), 409-424.

Wallace, D. S., Paulson, R. M., Lord, C. G., & Bond, C. F. (2005). Which behaviors do attitudes predict? Meta-analyzing the effects of social pressure and perceived difficulty. *Review of General Psychology*, 9, 214-227.

Woolley, S. L., Benjamin, W. J., & Woolley, A. W. (2004). Construct validity of a self-report measure of teacher beliefs related to student-centered and traditional approaches to teaching and learning. *Educational and Psychological Measurement,* 64, 319-331.

Yu, C. H. (2001). An introduction to computing and interpreting Cronbach Coefficient Alpha in SAS. *Proceedings of 26th SAS User Group International Conference*.

ACKNOWLEDGEMENTS

Dr. Anthony Coxon, Dr. Edith Neumann, Dr. Frank Gomez, and Mr. Henrique Santos